\documentclass[12pt]{article}
\usepackage{graphicx,hyperref,pifont}

\newfam\bboardfam
\font\tenbboard=msbm10  
 \font\sevenbboard=msbm7
   \font\fivebboard=msbm5 
\textfont\bboardfam=\tenbboard
\scriptfont\bboardfam=\sevenbboard
\scriptscriptfont\bboardfam=\fivebboard
\def\bboard{\fam\bboardfam\tenbboard}
\def\R{{\bboard R}}     

\newcommand{\st}{\mbox{$[S_{tot}]$}}

\title{Modular model building}

\author{Aneil Mallavarapu$^{1\ast}$, Matthew Thomson$^{1}$, \\ Benjamin Ullian$^1$, Jeremy Gunawardena,$^{1\ast}$
\\[0.5em]
\normalsize{$^{1}$Department of Systems Biology, Harvard Medical School}\\
\normalsize{200 Longwood Avenue, Boston, MA 02115, USA.}\\[0.2em]
\normalsize{$\ast$Corresponding authors:}\\
\normalsize{{\tt aneilbaboo@gmail.com} \hspace{0.2in} {\tt jeremy@hms.harvard.edu}} \\
\normalsize{Tel: {\sf (617) 432 4839}; Fax: {\sf (617) 432 5012}}
}
\date{}

\begin{document}

\maketitle

\begin{abstract}
Mathematical models are increasingly used in both academia and the pharmaceutical industry to understand how phenotypes emerge from systems of molecular interactions. However, their current construction as monolithic sets of equations presents a fundamental barrier to progress. Overcoming this requires modularity, enabling sub-systems to be specified independently and combined incrementally, and abstraction, enabling general properties to be specified independently of specific instances. These in turn require models to be represented as programs rather than as datatypes. Programmable modularity and abstraction enables libraries of modules to be created for generic biological processes, which can be instantiated and re-used repeatedly in different contexts with different components. We have developed a computational infrastructure to support this. We show here why these capabilities are needed, what is required to implement them and what can be accomplished with them that could not be done previously.
\end{abstract}

\clearpage

\section*{Introduction}

With the sequencing of the human genome, biological attention is shifting from characterizing molecular components to a systems-level understanding of how phenotypes emerge from molecular interactions \cite{kir05,wp04}. Mathematical models are increasingly used to shed light on this \cite{abls06,bfmhph,ldb05}. Such models present unusual challenges, not previously encountered in physics or engineering, upon which this paper focuses. 

Models may be static (eg: constraint based \cite{bfmhph}) or explicitly incorporate time. The latter are typically a form of dynamical system: they specify a set of molecular states and how those states evolve in time. Depending on the type of model, the states  may be discrete levels (boolean models \cite{laa06}), concentrations (ordinary differential equation models \cite{segm02}), molecular numbers (stochastic models \cite{rob05}) or sets of individual molecules (agent-based models \cite{epq04}), considered as functions of time or of space and time (spatial models \cite{sscl02}). While much of what follows may be broadly generalized, we focus here on dynamic models represented by ordinary differential equations,
\begin{equation}
\frac{dx}{dt} = f(x;a)
\label{e-ds}
\end{equation}
where $x \in \R^n$ is a vector of species concentrations, $a \in \R^m$ is a vector of parameter values and $f$ expresses the balance between the rates of production and consumption of each species. Such models are predominantly nonlinear. Although they may occasionally be analyzed mathematically \cite{gun05}, they are usually simulated numerically, for which parameter values must be specified and initial conditions chosen. 

Concern is sometimes raised about the complexity of such models and the large number of undetermined parameters. Model complexity is inevitable, however, if we wish to reflect the underlying biology \cite{segm02}, while methods for dealing with parameters are being developed \cite{abls06,rssm05}. Pharmaceutical companies like AstraZeneca and Pfizer as well as biotechnology companies like GNS and Entelos are now using modeling in drug development \cite{hmcydmgakd,bp03,hgld06}. Model complexity is not necessarily a barrier to usefulness and methodologies for constructing complex models in a controlled manner are therefore all the more important.

A variety of modeling tools are available \cite{ark01,rob05}, standards and ontologies formulated \cite{sbml,lfhbccchkmnssssw} and public model repositories established \cite{lbbcddlssssh}. With the right information to hand, it is straightforward to build a model. However, if model building is to be integrated into a research or development program, a single monolithic model is never sufficient. Simply to understand how a model works, it is often essential to create it incrementally, adding a few ingredients at a time and exploring the effect of alternative assumptions. More importantly, feedback between experiments and models leads to corrections or new assumptions, links appear to systems studied by others and new knowledge continually emerges in the literature. Such changes are readily incorporated into the mental ``models'' maintained by all biologists. Mathematical models lack such plasticity. Even simple biological changes can have profound effects on the equations, requiring new equations to be introduced or the modification of many existing equations and many terms in each. Beyond a certain level of model complexity, it is easier to build a new monolithic model from scratch. 

For example, Huang and Ferrell constructed an influential model of the MAP kinase cascade, which shed light on the decision making underlying maturation of Xenopus oocytes \cite{hf96}. Levchenko et al subsequently elucidated the surprising effect of a scaffold protein on MAP kinase signaling \cite{lbs00}. The second model contains exactly the same MAP kinase cascade as the first, based on identical assumptions, and differs only in the addition of one new component, the scaffold. Nevertheless, it was not obtained from the former by incremental extension; a new monolithic set of equations was constructed. Anyone wishing to build upon the work of Levchenko et al would have to do the same. The need to reinvent the wheel each time is a fundamental barrier to progress. 

Much of biology can be thought of as general processes, which operate on different components and which are combined in different ways. For example, all scaffolds behave in essentially the same way: they have no intrinsic enzymatic function but bind other proteins \cite{md03}. Scaffolds may differ in the number of binding partners and in the behavior of partners when bound but there is a core mechanism which must be incorporated in any model in which a scaffold participates. At present, such common behavior cannot be exploited in model construction. The corresponding equational details must be specified each time. Model construction would be far easier if these details could be specified once (by scaffolding experts, say) and this description re-used repeatedly by instantiating it with the particular binding partners and binding assumptions that are relevant to the context being modeled. Many other molecular processes are generic, in the sense that the same core mechanism is used with different components in different contexts. For instance, all receptor tyrosine kinase pathways are built from the following generic processes: receptor dimerization, endo- and exo-cytosis, endosomal recycling, multisite post-translational modification (including phosphorylation and ubiquitination), scaffolding, GTPase switching, membrane localization, nuclear import and export, etc. Model construction would be revolutionized if models could be built in a modular and incremental fashion from a library of expert descriptions of such generic processes. Models would then build upon each other, greatly increasing their scope and credibility. 

To accomplish this requires modularity, allowing sub-systems to be specified and then combined incrementally, and abstraction, allowing generic properties to be specified independently of specific instances. These in turn require models to be represented as programs within a computational infrastructure. Despite the variety of modeling tools available, none provide programmable modularity and abstraction. We have developed a novel computational infrastructure which does. It is open source and freely available and a detailed description appears elsewhere. Our purpose here is to show why modularity and abstraction are needed, what is required to implement them and what can be accomplished with them that could not be done previously.

\section*{Results}

\begin{figure}
\centering 
\includegraphics[viewport=14 14 582 504,width=\textwidth,height=\textheight,keepaspectratio]{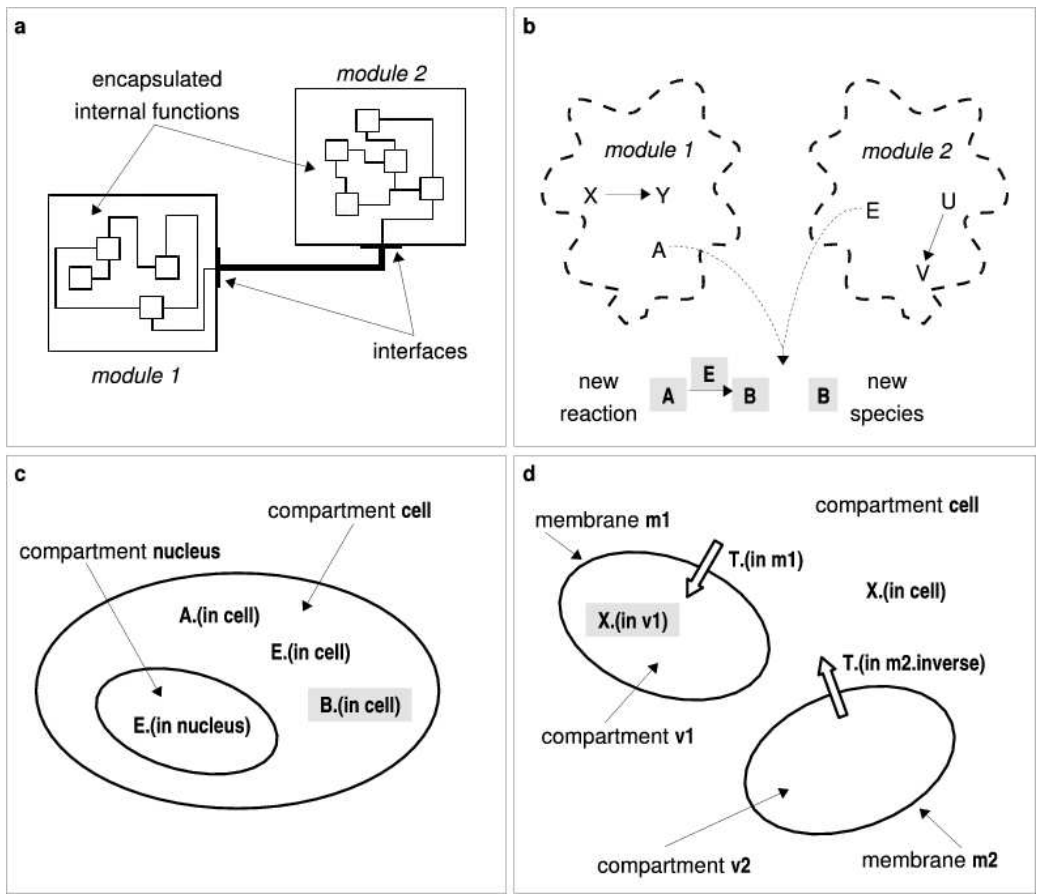}
\caption{Modularity, inference and identity.\label{f-mod}}
\end{figure}

{\noindent\bf Modularity and abstraction: models as programs\hspace{0.05in}} Modularity is a fundamental method for building complex engineering systems. Modules are sub-systems which are encapsulated to hide their internal complexity and inter-module communication is only permitted through specified interfaces (Figure~\ref{f-mod}a). By sub-dividing the design problem, modularity allows the engineer to subdue complexity. Although the same term is used in discussions of evolutionary mechanisms \cite{hhlm} we make no assumptions about evolution; modularity for us means a method of model construction. 

The advantages of the kind of ``engineering modularity'' just described have been previously appreciated and the capability is available in certain tools \cite{lhn04,gknrg,ww05}. However, encapsulating a module and specifying its interface at the time it is designed restrict the module's interactions to situations envisaged at design time. Biological modules have no natural encapsulation other than membranes: in the absence of physical separation, the components of one module may, in principle, always interact biochemically with the components of another. Our knowledge of which interactions actually occur may change over time in the light of new experiments. Whether or not such interactions are incorporated into a model may depend on the assumptions being made. In other words, not all interactions can be anticipated at design time. Furthermore, biological interaction creates entities, either new biochemical species or complexes of existing ones, which have to be represented in a model. This does not happen in engineering: wiring two transistors together does not create a capacitor. 

A flexible notion of modularity, adapted to biology, has to allow for the possibility that modules may be brought together in a way that was not envisaged at design time and that new entities (species, reactions) may arise from such modular composition (Figure~\ref{f-mod}b). The creation of new entities leads to two problems. First, the introduction of a new entity may have knock-on consequences in other modules, requiring yet more entities to be created, and these may in turn have further consequences and so on. Second, the new entity may already exist in some other module. If so, it is very important that a duplicate not be constructed.

To solve these problems a computational infrastructure needs, first, to be capable of reasoning over the modules and the properties of their components and inferring which entities need to be created (Figure~\ref{f-mod}b). In engineering modularity this reasoning is carried out by the designer at design time (Figure~\ref{f-mod}a). The capability needed here is analogous to the reasoning process used in artificial intelligence systems, where pattern-action rules, of the form $P \Rightarrow A$, specify actions ($A$) to be performed---including modification, removal or creation of entities---whenever there are entities matching the corresponding set of patterns ($P$). The ``many-pattern/many-entity'' matching problem is efficiently solved by the widely-used RETE algorithm \cite{forgy82}. Second, the infrastructure needs to give unique identities to its entities. This is a subtle problem because identities cannot simply be random tags: they must reflect the biological location of the entity and such locations can be complex entities in their own right (Figure~\ref{f-mod}c and d). 

To implement modularity and abstraction, models have to be represented as programs. Programming is needed to specify biological knowledge, to create new abstractions and to construct and reason about the new entities that arise from modular composition. This contrasts with SBML, the Systems Biology Markup Language \cite{sbml}, which treats models as datatypes and marks them up with meta-data, allowing a model created by one tool to be analyzed by another. SBML has made a vital contribution by nucleating the community of model builders and providing a de-facto standard for model communication \cite{lbbcddlssssh}. The next phase of model building may require a shift from datatypes to programs.

\begin{figure}
\centering 
\includegraphics[viewport=14 14 576 558,width=\textwidth,height=\textheight,keepaspectratio]{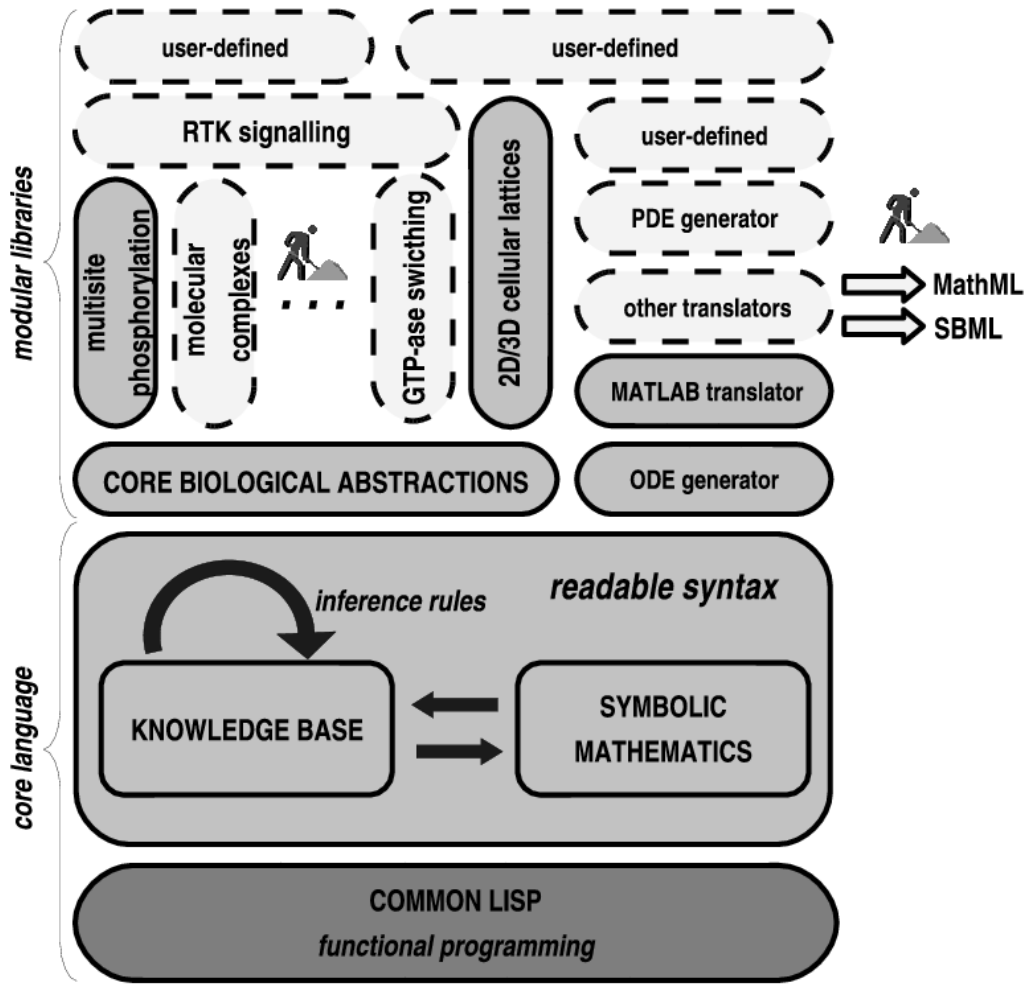}
\caption{Little b provides an extensible architecture.\label{f-aci}}
\end{figure}

We have designed a high-level programming language, {\em little b}, for modular model building. Biological knowledge specified as a program is evaluated by the {\em little b} computational infrastructure and translated into equations expressed in Matlab (Figure~\ref{f-aci}). Other simulation engines may be used instead and SBML may also be generated. {\em Little b} is implemented on top of Common Lisp \cite{graham96}, which was chosen because (1) Common Lisp is open source; (2) an open-source implementation of the RETE algorithm is available in the Lisp-based Intelligent Software Agents system (LISA); (3) Lisp's symbolic processing capability provides powerful mechanisms for implementing modularity and abstraction \cite{sm91}; (4) Lisp metaprograming allows {\em little b} to present a readable, biologically meaningful, non-Lisp-like syntax to the user (Figure~\ref{f-msp}d). 

{\noindent\bf Multisite phosphorylation as a generic module\hspace{0.05in}} Phosphorylation and dephosphorylation are ubiquitous and vital cellular regulatory mechanisms \cite{coh01}. Serine, threonine, tyrosine phosphorylation in eukaryotes usually takes place on multiple sites on a substrate \cite{coh00}. For instance, every component in the EGF signaling pathway, from the receptor to the transcription factors for immediately early genes, is phosphorylated several times. This makes multisite phosphorylation a prime candidate for a generic library module, which encodes current biological understanding and can be re-used repeatedly to make different models in different contexts.

The current understanding of multisite phosphorylation is typically complex (detailed citations are provided in the Supplementary Information). Two main features must be taken into account. Some kinases are known to be sequential, or nearly so, phosphorylating sites in a strict, or preferred, order, while others appear to phosphorylate sites randomly. Although site order has been less well-studied for phosphatases, they may exhibit similar preferences. Second, both enzymes may exhibit some degree of processivity, making several modifications in a single molecular encounter  \cite{hf96}. Degrees of processivity range from 1 modification per encounter (``distributivity''), for both kinases and phosphatases, to kinases which are ``highly processive'' on up to 20 sites. 

The following use cases emerge from this: kinase and phosphatase are both sequential but operate in reverse order, thereby maintaining a sequential set of $n+1$ phospho-forms (Figure~\ref{f-msp}c); kinase and phosphatase are both non-sequential with $2^n$ phospho-forms (Figure~\ref{f-msp}b); either enzyme may be processive to varying degree (Figure~\ref{f-msp}c has degree 2 for both enzymes), with distributivity corresponding to the limiting case of one modification at a time (Figure~\ref{f-msp}b). The program in Figure~\ref{f-msp}d uses a generic {\em little b} module that encompasses all the possibilities. This module can reproduce all mass-action multisite phosphorylation models known to us and can be incorporated in a modular way in descriptions of more complex systems.

We used the module in Figure~\ref{f-msp}d to study multistability in multisite phosphorylation. A substrate with $n$ sites can, in principle, exist in one of $2^n$ phospho-forms. We showed in recent work \cite{mg07} that, when both enzymes are distributive, there may be as many as $\lfloor(n+2)/2\rfloor$ stable distributions of phospho-forms at steady state ($\lfloor x\rfloor$ is the greatest integer not greater than $x$), suggesting the possibility for complex decision making. To reveal this multistability experimentally, we sought different starting conditions that would lead to different stable distributions. Although stable distributions can be analyzed mathematically, the dynamics leading to them can only be studied by simulation, which needs to be undertaken for different numbers of sites and sequential as well as non-sequential systems. The generic module in Figure~\ref{f-msp}d makes this straightforward. Figure~\ref{f-dyn} shows two scenarios in which appropriate starting conditions generate different stable phospho-form distributions. This behavior was representative of the multistable systems we studied. These predictions from simulation suggest a method for detecting multistability experimentally, which we are now testing in the laboratory. 

The multisite phosphorylation module shows that powerful abstractions for describing generic biological processes can be programmed from the core biological abstractions in {\em little b} (Figure~\ref{f-aci}). The next example shows that different modules can be combined together.

{\noindent\bf Developmental patterning on realistic cellular lattices\hspace{0.05in}} During initial patterning of the {\em Drosophila} embryo, maternal mRNAs stimulate expression of gap genes, followed by pair rule genes, followed by segment polarity genes, thereby establishing the anterior-posterior pattern of parasegments \cite{wol01}. The early stages of this process take place in the synctial blastoderm but the segment polarity genes turn on after cellularization. von Dassow et al created a computational infrastructure, Ingeneue \cite{mmov02}, for building models of gene regulatory networks in a 2D lattice of regular hexagonal cells and used it to build a model of {\em Drosophila} segment polarization \cite{vmmo00}. The proposed segment polarity network (Figure~\ref{f-network}b) was able to correctly stabilize a pre-pattern of Wingless and Engrailed expression over a wide range of different parameter values, suggesting that such robustness might be an evolutionary criterion for selecting network designs. This idea stimulated much interest \cite{wol01}.

In reality, {\em Drosophila} embryo epithelia are not regular hexagonal lattices of cells (Figure~\ref{f-lattice}a). Cells may be of different sizes and shapes and have different numbers of neighbors. Indeed, in proliferating animal epithelia, approximately 29\% of the cells resemble pentagons, 46\% resemble hexagons and 21\% resemble heptagons \cite{gpnp06}. The segment polarity network should produce the right patterning irrespective of which lattice emerges from cellularization. Determining whether the network in Figure~\ref{f-network} has this property would be a formidable challenge using traditional modeling tools. Yet it is an important question. Robustness to lattice variation is a more stringent requirement than robustness to parameter variation. The latter involves a mere numerical change in parameters without any change to the structure of the underlying equations; the former involves a restructuring of the equations themselves because the pattern and rates of cell-cell communication are altered. Robustness to lattice variation may thus exert greater selective pressure on network designs than does robustness to parameter variation.

Ingeneue provides computational support for building reactions in regular hexagonal cells, with cell-cell communication being modeled by interaction across apposed membrane segments. It keeps track of each apposed pair of membrane segments and all the molecular interactions across each apposition. The bookkeeping required is substantial but straightforward for a regular hexagonal lattice and such functionality is hardwired into Ingeneue. Each irregular lattice, however, needs its own bookkeeping scheme and the equations need to be rewritten to reflect each new scheme. Ingeneue does not have the capability to do this.

We implemented a generic cellular lattice module in {\em little b} for 2D polygonal (or 3D polyhedral) cells. The lattice module reads a list of coordinates of the vertices of the lattice, supplied by the user, and creates an internal representation of the cellular lattice. This module may be composed with any regulatory or protein interaction network module, using the same membrane apposition assumption for cell-cell communication as used in Ingeneue (Figure~\ref{f-network}). {\em Little b} works out all the bookkeeping. Modularity enables modeling of any gene regulatory network in any lattice of polygonal cells, subsuming and greatly extending the functionality of Ingeneue.

We used this module to examine the behavior of Figure~\ref{f-network} in four lattices, including a physiologically realistic lattice extracted from a {\em Drosophila} embryo image (Figure~\ref{f-lattice}a and b). The parameter robustness observed for the regular hexagonal lattice arises from a combination of ultrasensitive Hill functions and feedback \cite{vo02,ing04}. We chose two previously defined sets of parameter values \cite{vmmo00}, as described in Methods, and varied in one set the Hill function controlling an inter-cellular activation and in the other set an intra-cellular negative feedback loop (both labeled in Figure~\ref{f-network}b). The inter-cellular activation showed a range of parameter values for which all four lattices produced correct patterning (Figure~\ref{f-lattice}c), although the range was substantially smaller than for the hexagonal lattice alone. However, the intra-cellular negative feedback showed no parameter values for which all four lattices work correctly (Figure~\ref{f-lattice}d). Indeed, the {\em Drosophila} lattice never produces the correct pattern, despite a substantial range in which the regular hexagonal lattice works. The network in Figure~\ref{f-network}b seems to be highly sensitive to lattice geometry. 

It is conceivable that a relatively small change in the model will generate robust patterning for realistic lattices. Alternatively, significant control loops may be missing from our current understanding of this system. The point we wish to make in this paper is that {\em little b} allows exploration of these scientifically interesting questions, which were not possible to address before.

\section*{Discussion}

The multisite phosphorylation example shows that a generic biological process can be abstracted into the succinct module used in Figure~\ref{f-msp}d, while the Drosophila segmentation example shows that modules can be composed together. These capabilities lay the foundation for the vision of modular model building articulated in the Introduction. 

Many excellent modeling tools and methodologies are available. Some focus on particular biological processes, such as gene regulation \cite{rob05}, some favor specific biological domains, like immunology \cite{mxakjg} or neuroscience \cite{neuron,genesis}, some provide specialized capabilities for simulation \cite{bfgh04,lb05,mxakjg} or analysis \cite{xppaut}.

What makes {\em little b} different from existing tools is its programming language for modularity and abstraction. A language allows its user to describe novel situations not previously envisaged by the language designer. Most tools support model building through some form of ``template'', restricting the user to what is provided in the menu of available templates. While templates are implemented in a programming language, the language itself is not accessible to the user. Templates cannot provide the functionality of Figure~\ref{f-msp}d, nor can the user construct new templates. In contrast, rule-based methodologies (rules here being different from pattern-action rules) include languages for model building, which provide powerful capabilities for representing protein domain interactions and tracking the combinatorics of molecular complexes \cite{ccdfs04,lb05,hfbphf}. However, these languages lack control and data structures commonplace in high-level programming languages. Figure~\ref{f-msp}d might conceivably be written in some of them but the effort required would be prohibitive. Cellerator \cite{slmwm} provides templates for model building through Mathematica. The functionality of Figure~\ref{f-msp}d could be implemented in this or other high-level programming languages. Such a program, however, would solve only one isolated problem---that of multisite phosphorylation. It could not be composed with other similar programs in a modular way, as the lattice module can be composed with any regulatory network module. To do this would require also implementing the capabilities for modularity and abstraction (identity, pattern-action rules, etc) provided by {\em little b}.

Graphical languages have also been proposed for biological specification \cite{kfmo05,kawp06}. Experience in engineering design suggests that while they are excellent for describing structure, such as the layout of an integrated circuit, function is best described through textual languages (Verilog, VHDL). Accordingly, we anticipate that interfaces based on graphical languages will take advantage of textual languages ``under the hood'', much as modern tools for constructing web pages are user-friendly vehicles for generating HTML.

We have focused in this paper on the transition from monolithic to modular models and on the computational infrastructure needed to support this. Not only is this required to build models more effectively, it is essential for their credibility. Models, particularly complex ones, are usually published as supplementary information. Even the most conscientious reviewer is unlikely to be able to subject such a model to the same level of scrutiny as a published experimental method or mathematical proof. Models may sometimes be submitted to a public repository but few others are likely to want to use an existing model without also wanting to change it, with all the attendant difficulties noted previously. Accordingly, monolithic models may have been closely studied only by their creators, a situation of some concern in an emerging discipline. Modular models, in contrast, can be pulled apart and their component modules evaluated, modified and recombined. Generic library modules, such as that for multisite phosphorylation, could be developed and refined by experts and made available to all model builders, thereby creating a scientific ``commons'' for model building. {\em Little b} allows any user to develop novel abstractions of the biology being studied and to contribute these back to the community, thereby allowing the field to evolve in a decentralized manner that enables us to build upon each other's work rather than having to recreate it. The models that result from this may be more complex but their credibility, reliability and usefulness will be more easily established. Modular model building will provide a more robust foundation for systems biology.

\section*{Methods}
\small
{\noindent\bf Computational infrastructure\hspace{0.05in}} {\em Little b} was developed within the LispWorks environment (LispWorks Ltd, Cambridge, UK). It is freely and publicly available from {\small\sf vcp.med.harvard.edu}, {\small\sf littleb.org} or \mbox{{\small\sf sourceforge.net}}. The computational infrastructure compiles a biological description expressed as a {\em little b} program into Matlab (The MathWorks, Natick, MA) files. Rate equations are either automatically derived using mass-action assumptions or the user can provide phenomenological rate functions (for instance, the Hill functions used in the segment polarity models). The symbolic mathematics subsystem (Figure~\ref{f-aci}) can accommodate rational functions of several variables, with arbitrary real exponents. Dimensions and units are consistently handled. The correctness of the infrastructure was tested by construction of a series of examples of increasing complexity, including four previously developed models \cite{hf96,lbs00,vmmo00,bi99}. In addition to reproducing Matlab results, the equations and their internal representations were checked. Lisp evaluation times range from under 1 second for the multisite phosphorylation module in Figure~\ref{f-msp}e to 11 minutes for the segment polarity module on the {\em Drosophila} lattice, which has 104 cells, 3439 species and 13328 reactions. (Timings on an IBM T43p laptop, Pentium M, 2.1GHz, 1Gb RAM).

\vspace{0.1in}
{\noindent\bf Segment polarization\hspace{0.05in}} Lattices were generated in Matlab by choosing a set of points and using a Voronoi tessellation to produce polygonal cells. For the {\em Drosophila} lattice, the points were selected manually as the centers of the biological cells in the embryo photograph, as in Figure~\ref{f-lattice}a. A Matlab script generates the $(x,y)$ coordinates of the vertices of each cell, along with the cell areas and the lengths of the apposed membrane segments. The generic cellular lattice module then uses this data to construct the resulting compartments and membranes. Figure~\ref{f-lattice}b shows the four lattices used for this study, on which are superimposed the pre-pattern (initial condition) of cells in which the levels of Wingless mRNA (wg) and protein (WG) and Engrailed mRNA (en) and protein (EN) are set to normalized concentrations of 1, as previously \cite{vmmo00}. All other components are initially zero, with the exception of the basal activator of cid expression, which has normalized concentration of 0.4 in each cell, as previously \cite{vmmo00}. The four lattices are: {\bf Hexagonal}, corresponding to the regular hexagonal lattice used previously \cite{vmmo00}; {\bf Drosophila}, extracted from the embryo photograph; {\bf Rectangular}, in which the cells are rectangular but come in two sizes, arranged in alternating columns; {\bf Shifted rectangular}, in which the lattice is identical to the rectangular lattice but the pre-pattern is shifted to the right.

We used identical assumptions to von Dassow {\em et al} to represent the regulatory network in Figure~\ref{f-network}a, forgoing later modifications \cite{ing04}. We did not wrap lattices onto a torus \cite{vmmo00}, as such double periodicity makes no sense for irregular lattices. We checked for edge effects by embedding one lattice inside a larger one; we found no evidence for major changes in behavior. We chose two previously used parameter sets \cite{vmmo00} but found that the high Hill coefficients gave rise to unphysiological oscillations in some components (Supplementary Figure~2). We were able to find lower Hill coefficients without jeopardizing correct segmentation on the hexagonal lattice. The parameter values used in Figure~\ref{f-lattice}c were derived in this way from the Yippee parameter set, while those in Figure~\ref{f-lattice}d were derived from parameter set four, as detailed in Supplementary Table~1. In running the simulations we found occasional slow decays beyond the 1000 minutes used previously \cite{vmmo00} (Supplementary Figure~2). We therefore scored correct segmentation by simulating for 5000 simulated minutes, thresholding Wingless and Engrailed values as previously \cite{vmmo00}, and checking if the results agreed with the pre-pattern. Edge cells were ignored in scoring. 

\section*{Acknowlegements}

We are grateful to the Bauer Center for Genomics Research and to Craig Muir for support during the initial phase of this work. We thank, especially, Steve Harrison, Ed Harlow, Marc Kirschner and Rebecca Ward of the Harvard Medical School for enabling this work to come to fruition through their support for the Virtual Cell Program. We thank Peter Lawrence for the {\em Drosophila} image in Figure~\ref{f-lattice}a, Radhika Nagpal for access to material in press, David Young for his open-source implementation of LISA and Dave Fox of LispWorks for his assistance with the Lisp environment. We thank Carl Pabo, Brian Seed and Rebecca Ward for their insightful comments and the other members of the Virtual Cell Program for their assistance.


\bibliographystyle{plain}
\bibliography{/home/jhcg/work/BIB/bio}

\begin{thebibliography}{10}

\bibitem{abls06}
B.~B. Aldridge, J.~M. Burke, D.~A. Lauffenburger, and P.~K. Sorger.
\newblock Physicochemical modelling of cell signalling pathways.
\newblock {\em Nat. Cell Biol.}, 8:1195--203, 2006.

\bibitem{ark01}
A.~P. Arkin.
\newblock Synthetic cell biology.
\newblock {\em Curr. Opin. Biotechnol.}, 12:638--44, 2001.

\bibitem{bp03}
A.~L. Bangs and T.~S. Paterson.
\newblock Finding value in in-silico biology.
\newblock {\em Biosilico}, 1:18--22, 2003.

\bibitem{bfmhph}
S.~A. Becker, A.~M. Feist, M.~L. Mo, G.~Hannum, B.~O. Palsson, and M.~J.
  Herrgard.
\newblock Quantitative prediction of cellular metabolism with constraint-based
  models: the {COBRA} {T}oolbox.
\newblock {\em Nat. Protoc.}, 2:727--38, 2007.

\bibitem{bi99}
U.~Bhalla and R.~Iyengar.
\newblock Emergent properties of networks of biological signalling pathways.
\newblock {\em Science}, 283:381--7, 1999.

\bibitem{bfgh04}
M.~L. Blinov, J.~R. Faeder, B.~Goldstein, and W.~S. Hlavacek.
\newblock {BioNetGen}: software for rule-based modeling of signal transduction
  based on the interactions of molecular domains.
\newblock {\em Bioinformatics}, 20:3289--91, 2004.

\bibitem{genesis}
J.~M. Bower and D.~Beeman.
\newblock {\em The Book of GENESIS: Exploring Realistic Neural Models with the
  GEneral NEural SImulation System}.
\newblock Springer, New York, USA, 1998.

\bibitem{neuron}
N.~T. Carnevale and M.~L. Hines.
\newblock {\em The NEURON Book}.
\newblock Cambridge University Press, Cambridge, UK, 2006.

\bibitem{ccdfs04}
N.~Chabrier-Rivier, M.~Chiaverini, V.~Danos, F.~Fages, and V.~Sch{\"a}cter.
\newblock Modeling and querying biomolecular interaction networks.
\newblock {\em Theor. Comput. Sci.}, 325:25--44, 2004.

\bibitem{coh00}
P.~Cohen.
\newblock The regulation of protein function by multisite phosphorylation - a
  25 year update.
\newblock {\em Trends Biochem. Sci.}, 25:596--601, 2000.

\bibitem{coh01}
P.~Cohen.
\newblock The role of reversible protein phosphorylation in health and disease.
\newblock {\em Eur. J. Biochem.}, 268:5001--10, 2001.

\bibitem{xppaut}
B.~Ermentrout.
\newblock {\em Simulating, Analyzing, and Animating Dynamical Systems: A Guide
  to Xppaut for Researchers and Students}.
\newblock SIAM, 2002.

\bibitem{epq04}
D.~D. Errampalli, C.~Priami, and P.~Quaglia.
\newblock A formal language for computational systems biology.
\newblock {\em OMICS}, 8:370--80, 2004.

\bibitem{forgy82}
C.~L. Forgy.
\newblock Rete: a fast algorithm for the many pattern/many object pattern match
  problem.
\newblock {\em Artificial Intelligence}, 19:17--37, 1982.

\bibitem{gpnp06}
M.~C. Gibson, A.~B. Patel, R.~Nagpal, and N.~Perrimon.
\newblock The emergence of geometric order in proliferating metazoan epithelia.
\newblock {\em Nature}, 442:1038--41, 2006.

\bibitem{gknrg}
A.~Ginkel, A.~Kremling, T.~Nutsch, R.~Rehner, and E.~D. Gilles.
\newblock Modular modelling of cellular systems with {ProMot/Diva}.
\newblock {\em Bioinformatics}, 19:1169--76, 2003.

\bibitem{graham96}
P.~Graham.
\newblock {\em ANSI Common Lisp}.
\newblock Series in Artificial Ingelligence. Prentice Hall, 1996.

\bibitem{gun05}
J.~Gunawardena.
\newblock Multisite protein phosphorylation makes a good threshold but can be a
  poor switch.
\newblock {\em Proc. Natl. Acad. Sci. USA}, 102:14617--22, 2005.

\bibitem{hmcydmgakd}
T.~Haberichter, B.~M{\"a}dge, R.~A. Christopher, N.~Yoshioka, A.~Dhiman,
  R.~Miller, R.~Gendelman, S.~V. Aksenov, I.~G. Khalil, and S.~F. Dowdy.
\newblock A systems biology dynamical model of mammalian {G1} cell cycle
  progression.
\newblock {\em Mol. Syst. Biol.}, 3:84, 2007.

\bibitem{hhlm}
L.~H. Hartwell, J.~J. Hopfield, S.~Leibler, and A.~W. Murray.
\newblock From molecular to modular cell biology.
\newblock {\em Nature}, 402:C47--52, 1999.

\bibitem{hgld06}
B.~S. Hendriks, G.~J. Griffiths, R.~Benson, D.~Kenyon, M.~Lazzara, J.~Swinton,
  S.~Beck, M.~Hickinson, J.~M. Beusmans, D.~Lauffenburger, and D.~de~Graaf.
\newblock Decreased internalisation of {ErbB1} mutants in lung cancer is linked
  with a mechanism conferring sensitivity to gefitinib.
\newblock {\em IEE Proc. Syst. Biol.}, 153:457--66, 2006.

\bibitem{hfbphf}
W.~S. Hlavacek, J.~R. Faeder, M.~L. Blinov, R.~G. Posner, M.~Hucka, and
  W.~Fontana.
\newblock Rules for modeling signal-transduction systems.
\newblock {\em Sci. STKE}, 344:re6, 2006.

\bibitem{hf96}
C.-Y.~F. Huang and J.~E. Ferrell.
\newblock Ultrasensitivity in the mitogen-activated protein kinase cascade.
\newblock {\em Proc. Natl. Acad. Sci. USA}, 93:10078--83, 1996.

\bibitem{sbml}
M.~Hucka, A.~Finney, H.~Bolouri, J.~C. Doyle, and H.~Kitano.
\newblock The {S}ystems {B}iology {M}arkup {L}anguage ({SBML}): a medium for
  representation and exchange of biochemical network models.
\newblock {\em Bioinformatics}, 19:524--31, 2003.

\bibitem{ing04}
N.~T. Ingolia.
\newblock Topology and robustness in the {{\em Drosophila}} segment polarity
  network.
\newblock {\em PLoS Biology}, 2:0805--15, 2004.

\bibitem{kir05}
M.~Kirschner.
\newblock The meaning of systems biology.
\newblock {\em Cell}, 121:503--4, 2005.

\bibitem{kfmo05}
H.~Kitano, A.~Funahashi, Y.~Matsuoka, and K.~Oda.
\newblock Using process diagrams for the graphical representation of biological
  networks.
\newblock {\em Nat. Biotechnol.}, 23:961--6, 2005.

\bibitem{kawp06}
K.~W. Kohn, M.~I. Aladjem, J.~N. Weinstein, and Y.~Pommier.
\newblock Molecular interaction maps of bioregulatory networks: a general
  rubric for systems biology.
\newblock {\em Mol. Biol. Cell}, 17:1--13, 2006.

\bibitem{lbs00}
A.~Levchenko, J.~Bruck, and P.~W. Sternberg.
\newblock Scaffold proteins may biphasically affect the levels of
  mitogen-activated protein kinase signaling and reduce its threshold
  properties.
\newblock {\em Proc. Natl. Acad. Sci.}, 97:5818--23, 2000.

\bibitem{laa06}
S.~Li, S.~M. Assmann, and R\'eka Albert.
\newblock Predicting essential components of signal transduction networks: a
  dynamic model of guard cell abscissic acid signaling.
\newblock {\em PLoS Biol.}, 4:e312, 2006.

\bibitem{lhn04}
C.~M. Lloyd, M.~D.~B. Halstead, and P.~F. Nielsen.
\newblock {CellML}: its future, present and past.
\newblock {\em Prog. Biophys. Mol. Biol.}, 85:433--50, 2004.

\bibitem{lb05}
L.~Lok and R.~Brent.
\newblock Automatic generation of cellular reaction networks with {M}oleculizer
  1.0.
\newblock {\em Nat. Biotechnol.}, 23:131--6, 2005.

\bibitem{ldb05}
W.~J.~R. Longabaugh, E.~H. Davidson, and H.~Bolouri.
\newblock Computational representation of developmental genetic regulatory
  networks.
\newblock {\em Dev. Biol.}, 283:1--16, 2005.

\bibitem{mxakjg}
M.~Meier-Schellersheim, X.~Xu, B.~Angermann, E.~J. Kunkel, T.~Jin, and R.~N.
  Germain.
\newblock Key role of local regulation in chemosensing revealed by a new
  molecular interaction-based modeling method.
\newblock {\em PloS Comput. Biol.}, 2:0710--24, 2006.

\bibitem{mmov02}
E.~Meir, E.~M. Munro, G.~M. Odell, and G.~von Dassow.
\newblock Ingeneue: a versatile tool for reconstituting genetic networks, with
  examples from the segment polarity network.
\newblock {\em J. Exp. Zoolog. Part B}, 294:216--51, 2002.

\bibitem{md03}
D.~K. Morrison and R.~J. Davis.
\newblock Regulation of {MAP} kinase signalling modules by scaffold proteins in
  mammals.
\newblock {\em Annu. Rev. Cell Dev. Biol.}, 19:91--118, 2003.

\bibitem{lbbcddlssssh}
N.~Le Nov{\'e}re, B.~Bornstein, A.~Broicher, M.~Courtot, M.~Donizelli,
  H.~Dharuri, L.~Li, H.~Sauro, M.~Schilstra, J.~L. Snoep, H.~D. Spence, and
  M.~Hucka.
\newblock Biomodels database: a free, centralized database of curated,
  published, quantitative kinetic models of biochemical and cellular systems.
\newblock {\em Nucleic Acids Res.}, 34:D689--91, 2006.

\bibitem{lfhbccchkmnssssw}
N.~Le Nov{\'e}re, A.~Finney, M.~Hucka, U.~S. Bhalla, F.~Compagne,
  J.~Collado-Vides, E.~J. Crampin, M.~Halstead, E.~Klipp, P.~Mendes,
  P.~Nielsen, H.~Sauro, B.~Shapiro, J.~L. Snoep, H.~D. Spence, and B.~L.
  Wanner.
\newblock Minimum information requested in the annotation of biochemical models
  ({MIRIAM}).
\newblock {\em Nat. Biotechnol.}, 23:1509--15, 2005.

\bibitem{rob05}
S.~Ramsey, D.~Orrell, and H.~Bolouri.
\newblock {DIZZY}: stochastic simulation of large-scale genetic regulatory
  networks.
\newblock {\em J. Bioinform. Comput. Biol.}, 3:415--36, 2005.

\bibitem{rssm05}
D.~A. Rand, B.~V. Shulgin, J.~D. Salazar, and A.~J. Millar.
\newblock Uncovering the design principles of circadian clocks: mathematical
  analysis of flexibility and evolutionary goals.
\newblock {\em J. Theor. Biol.}, 238:616--35, 2005.

\bibitem{segm02}
B.~Schoeberl, C.~Eichler-Jonsson, E.~D. Gilles, and G.~M{\"u}ller.
\newblock Computational modeling of the dynamics of the {MAP} kinase cascade
  activated by surface and internalized {EGF} receptors.
\newblock {\em Nat. Biotechnol.}, 20:370--5, 2002.

\bibitem{slmwm}
B.~E. Shapiro, A.~Levchenko, E.~M. Meyerowitz, B.~J. Wold, and E.~D. Mjolsness.
\newblock Cellerator: extending a computer algebra system to include
  biochemical arrows for signal transduction simulations.
\newblock {\em Bioinformatics}, 19:677--8, 2003.

\bibitem{sm91}
K.~H. Sinclair and D.~A. Moon.
\newblock The philosophy of {Lisp}.
\newblock {\em Commun. ACM}, 34:41--7, 1991.

\bibitem{sscl02}
B.~M. Slepchenko, J.~C. Schaff, J.~H. Carson, and L.~M. Loew.
\newblock Computational cell biology: spatiotemporal simulation of cellular
  events.
\newblock {\em Annu. Rev. Biophys. Biomol. Struct}, 31:423--41, 2002.

\bibitem{mg07}
M.~Thomson and J.~Gunawardena.
\newblock A systems model of multisite phosphorylation reveals a capability for
  complex decision making.
\newblock Submitted, 2007.

\bibitem{vmmo00}
G.~von Dassow, E.~Meir, E.~M. Munro, and G.~M. Odell.
\newblock The segment polarity network is a robust developmental module.
\newblock {\em Nature}, 406:188--92, 2000.

\bibitem{vo02}
G.~von Dassow and G.~M. Odell.
\newblock Design and constraints of the {Drosophila} segment polarity module:
  robust spatial patterning emerges from intertwined cell state switches.
\newblock {\em J. Exp. Zoolog. Part B}, 294:179--215, 2002.

\bibitem{ww05}
K.~Webb and T.~White.
\newblock {UML} as a cell and biochemistry modelling language.
\newblock {\em BioSystems}, 80:283--302, 2005.

\bibitem{wp04}
H.~V. Westerhoff and B.~O. Palsson.
\newblock The evolution of molecular biology into systems biology.
\newblock {\em Nat. Biotechnol.}, 22:1249--52, 2004.

\bibitem{wol01}
L.~Wolpert.
\newblock {\em Principles of Development}.
\newblock Oxford University Press, 2001.

\end{thebibliography}

\section*{Figure Legends}

{\noindent\bf Figure 1} Modularity, inference and identity. {\bf a} Engineering modularity exposes restricted functionality through interfaces, while hiding internal complexity behind barriers. {\bf b} Modularity for biological models must allow for the possibility that any molecule may interact with any other molecule. $E$ is an enzyme which converts $A$ to $B$. Module 1 contains $A$ but not $E$ or $B$ while module 2 contains $E$ but not $A$ or $B$. Composing modules 1 and 2 results in a new reaction, $A \rightarrow B$, and a new species, $B$, not previously present in either module. A module must work correctly in contexts determined by other modules whose characteristics are not known in advance of module composition. {\em Little b's} computational infrastructure uses reasoning to infer the presence of the highlighted entities. {\bf c} Unique identities must encode information about location. $E$ converts $A$ to $B$ and is present in both the {\em cell} compartment and the {\em nucleus} compartment. $A$, however, is only present in the {\em cell}. The system should infer that $B$ is present in the {\em cell} but, in the absence of other information, should not infer that it is present in the {\em nucleus}. {\bf d} Membranes are complex locations. $T$ transports $X$ uni-directionally across a membrane. $T$ is oriented in membrane $m1$ of vesicle $v1$ to transport $X$ into $v1$ but is oppositely oriented in membrane $m2$ of vesicle $v2$. If $X$ is present in the {\em cell} compartment then the computational infrastructure should infer that it is in $v1$ but, in the absence of other information, should not infer that it is in $v2$. Membranes encode information about their two adjacent volume compartments and molecules are oriented by locating them in either the standard (default) membrane or its {\em inverse}. The behavior of $T$ and $X$ is described once but works correctly irrespective of $T$'s location.

\vspace{0.1in}
{\noindent\bf Figure 2} Little b provides an extensible architecture, permitting development of new generic modules.  The core language extends Common Lisp with new syntax, a reasoning system and symbolic mathematics. Modular libraries provide both biological and mathematical abstractions in a hierarchical fashion. A library of core biological abstractions defines reusable constructs for representing and reasoning about reactions, molecular complexes and biochemical locations. Higher-order modules, such as ``multisite phosphorylation'' and ``2D/3D cellular lattices'' discussed in the text, can be programmed on top of the core abstractions. Users have access to all levels of the hierarchy and can build new modules which extend the biological or mathematical capabilities. Yellow, dashed boxes indicate libraries that are envisaged or under development, while pink, full boxes show the currently implemented little b computational infrastructure.

\vspace{0.1in}
{\noindent\bf Figure 3} Reaction schemes for multisite phosphorylation. {\bf a} Processive enzymatic phosphorylation and dephosphorylation assumes a single enzyme-substrate complex and irreversible release of products, $P_1, \cdots, P_m$. Distributivity corresponds to $m = 1$. ATP and ADP are assumed held constant by mechanisms not explicitly represented. The kinetics are determined by mass-action. {\bf b} Distributive, non-sequential phosphorylation and dephosphorylation with $n = 2$ sites. Phospho-forms are denoted $S_b$ where $b$ is a sequence of $n$ bits (0 or 1). $E$ kinase, $F$ phosphatase, $S$ substrate. {\bf c} Processive, sequential phosphorylation and dephosphorylation with $n = 4$ and processivity $2$. Enzyme release steps are omitted for clarity. {\bf d} Example program using the generic module (lines 4/5 and 6/7) for multisite phosphorylation, which can generate any reaction scheme like {\bf b, c}. Bold blue text, object classes; bold italic pink text, keywords. The module is instantiated for {\bf c} but can be instantiated for {\bf b} by changing the number of sites to $2$ (line 3), the processivity to $1$ and the mode to non-sequential (lines 5/7). Model equations are generated after rate constants and initial conditions are specified (not shown).

\begin{figure}
\centering 
\includegraphics[viewport=14 14 570 656,width=\textwidth,height=\textheight,keepaspectratio]{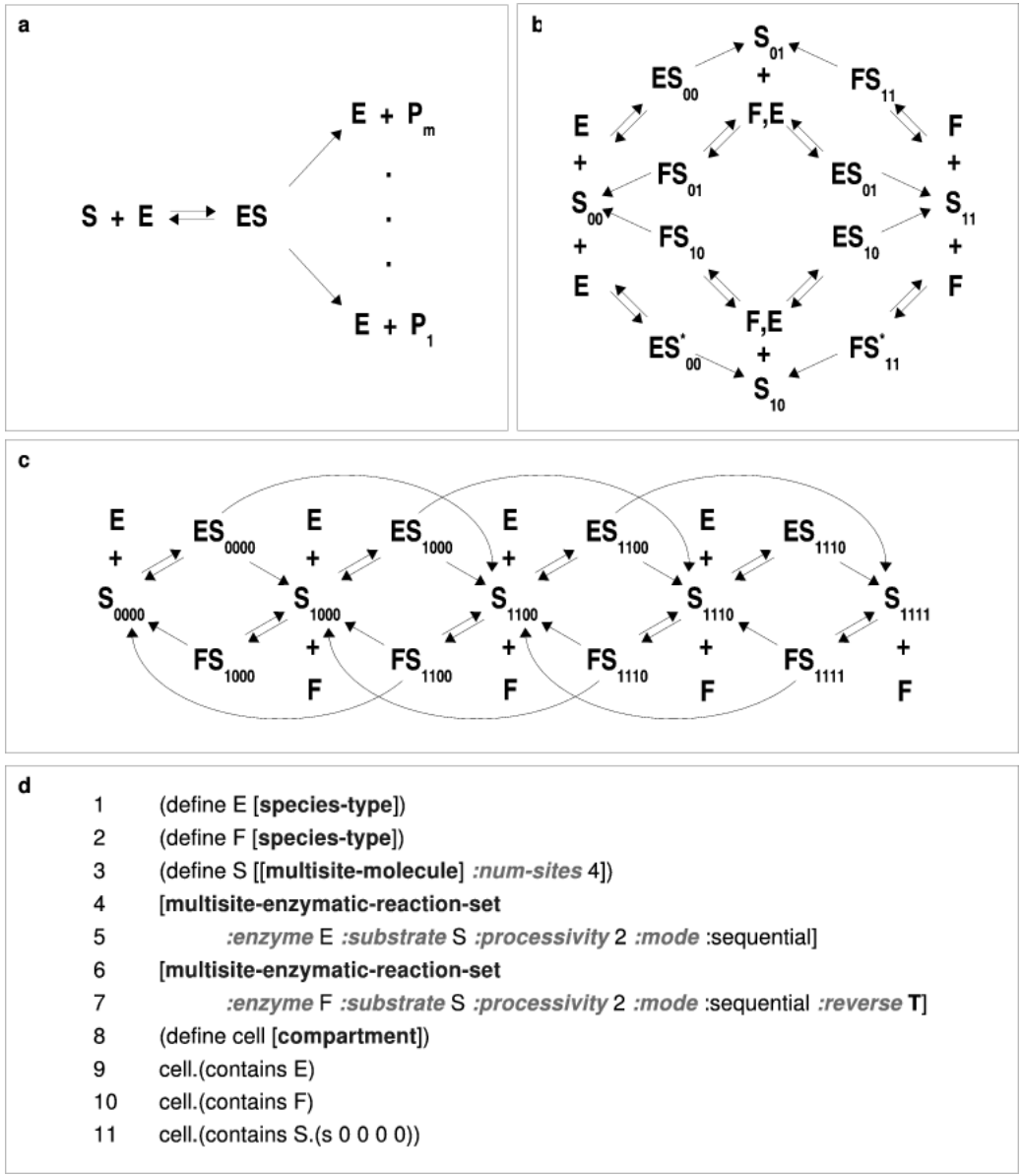}
\caption{Reaction schemes for multisite phosphorylation.\label{f-msp}}
\end{figure}

\begin{figure}
\centering 
\includegraphics[viewport=14 14 517 794,width=\textwidth,height=0.9\textheight,keepaspectratio]{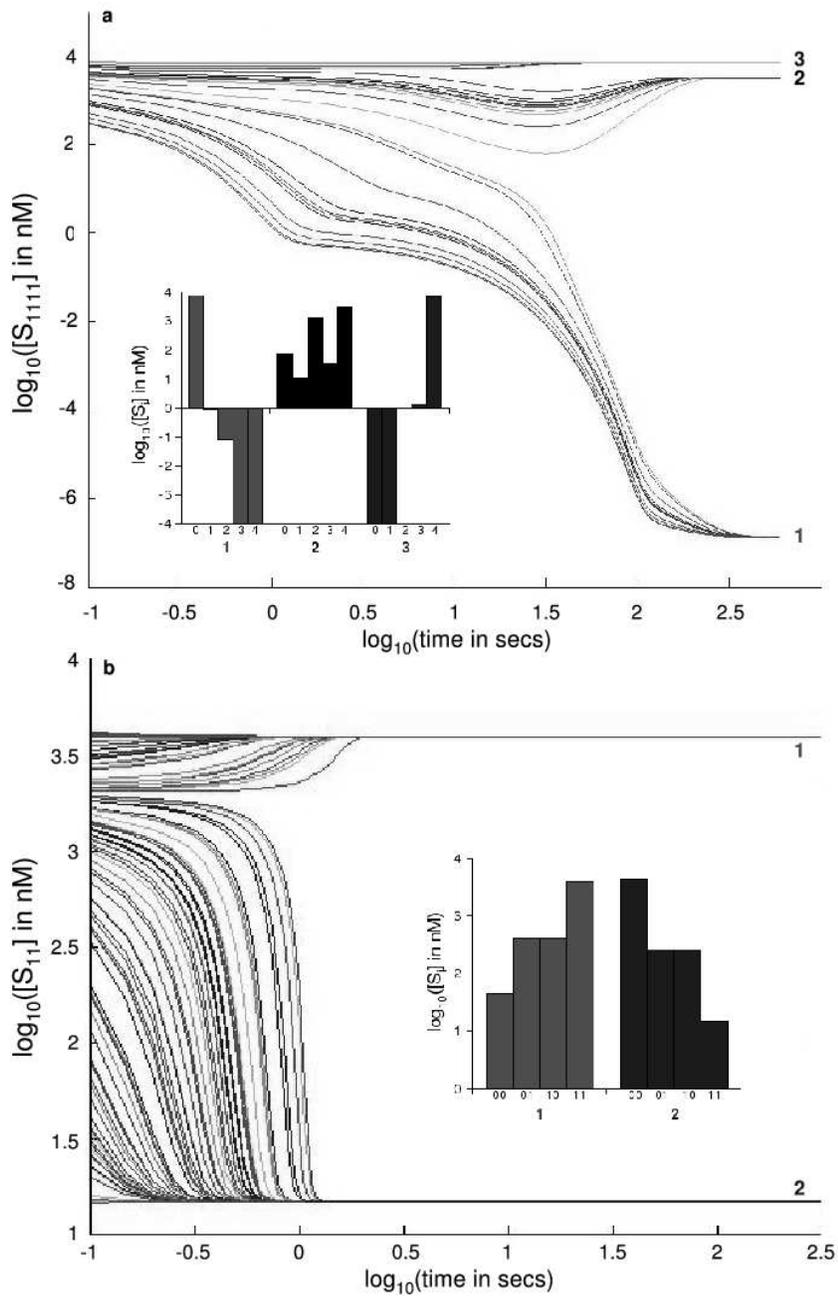}
\caption{Multistability in multisite phosphorylation.\label{f-dyn}}
\end{figure}

\vspace{0.1in}
{\noindent\bf Figure 4} Multistability in multisite phosphorylation. Rate constants and initial conditions are given in Supplementary Figure~1. {\bf a} Distributive, sequential phosphorylation and dephosphorylation, with $n = 4$, as Figure~\ref{f-msp}c but with $k = 1$. Substrate is initially present as $[S_{0000}] = \alpha\st$, $[S_{1111}] = (1 - \alpha)\st$, where $\alpha$ is drawn randomly from the uniform distribution on $[0,1]$ and $\st$ is the total amount of substrate present. $[-]$ denotes concentration. Vertical axis, concentration of $S_{1111}$; horizontal axis, time; $\log_{10}$ scales on both. The initial conditions find the three stable phospho-form distributions shown in the insert, for appropriate values of $\alpha$. In the inset, phospho-forms are designated $0,1,2,3,4$ by number of phosphorylations. {\bf b} Distributive, non-sequential phosphorylation and dephosphorylation, as Figure~\ref{f-msp}b. Initial substrate is a random combination of $S_{00}$ and $S_{11}$, as previously, leading to the two stable phospho-form distributions shown in the inset. Vertical axis, concentration of $S_{11}$; horizontal axis, time; $\log_{10}$ scales on both.

\vspace{0.1in}
{\noindent\bf Figure 5} Modular construction of developmental networks in arbitrary cellular lattices. {\bf a} Polygonal lattice of cells. The user provides the vertex coordinates to {\em little b's} generic lattice module, which creates an internal representation of the lattice. {\bf b} The segment polarity gene regulation network after von Dassow {\em et al} \cite{vmmo00}. Positive feedback of Wingless protein (WG) on its mRNA (wg) and repression of Engrailed mRNA (en) by cleaved Cubitus Interruptus (CN) are both included. Labels show the interactions ($c$ = WGen and $c$ = PTCCID) which are varied in Figure~\ref{f-lattice}. {\em Little b} can take any polygonal lattice and any network of reactions and put the two together in a modular way. Each cell acquires the network and two cells interact across their common membrane segment using the same mechanism as in von Dassow et al \cite{vmmo00}. Each cell and each membrane apposition is a potential location for species or reactions, which must all be individually accounted for to construct the equations.

\begin{figure}
\centering 
\includegraphics[viewport=14 14 566 231,width=\textwidth,height=\textheight,keepaspectratio]{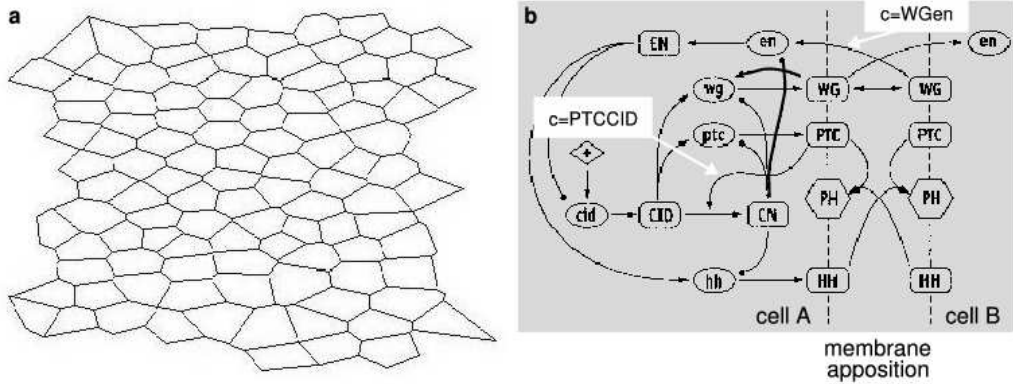}
\caption{Modular construction in arbitrary cellular lattice.\label{f-network}}
\end{figure}

\begin{figure}
\centering 
\includegraphics[viewport=14 14 592 434,width=\textwidth,height=\textheight,keepaspectratio]{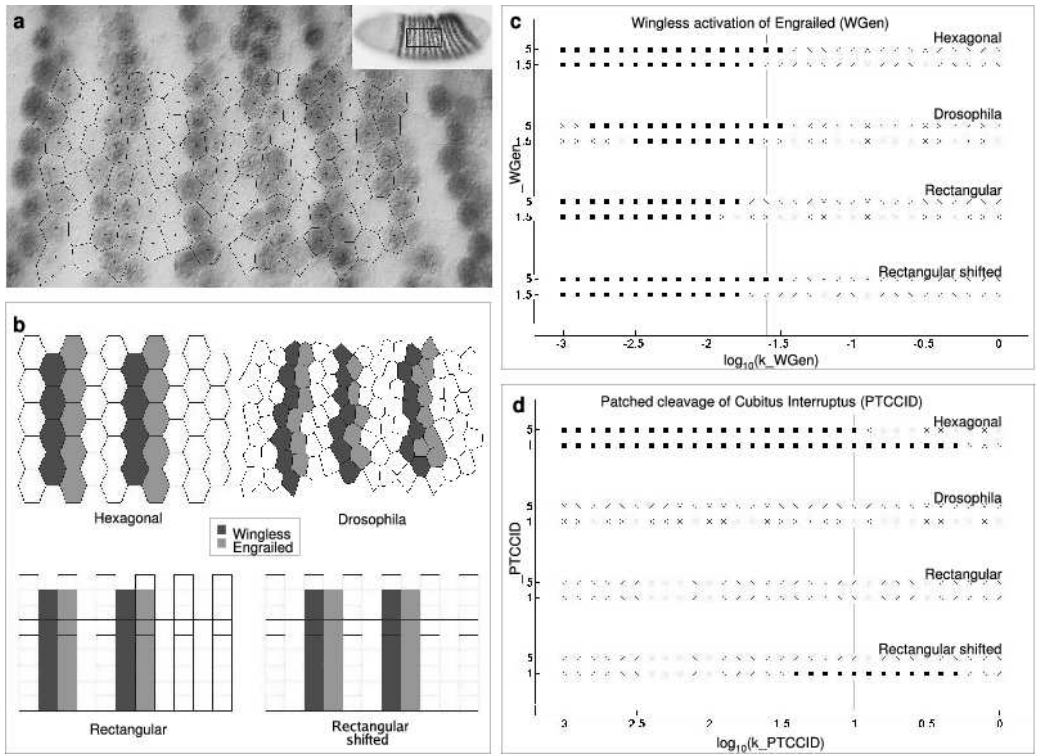} 
\caption{Segment polarization in different cellular lattices.\label{f-lattice}}
\end{figure}

\vspace{0.1in}
{\noindent\bf Figure 6} Segment polarization in different cellular lattices. {\bf a} Image of a {\em Drosophila} embryo, with the extracted cellular lattice superimposed. {\bf b} Four lattices showing the pre-pattern where Wingless (red) and Engrailed (green) are high. For correct segmentation, the regulatory network must stabilize this pre-pattern, starting from the pre-pattern as initial condition. {\bf c and d} Correct (\ding{110}) or incorrect (\ding{53}) segmentation for the lattices listed on the right. The half-maximal value, $k\_c$ (horizontal axis), and the Hill coefficient, $\nu\_c$ (vertical axis), of a Hill function, $x^{\nu\_c}/((k\_c)^{\nu\_c} + x^{\nu\_c})$, describing one of the connections, $c$, in Figure~\ref{f-network}b, are varied. The half-maximal value varies horizontally on a $\log_{10}$ scale, while the Hill coefficient takes either a low (1 or 1.5) or high (5) value. The parameter values other than $\nu_c$ and $k_c$ are obtained from two previously defined parameter sets \cite{vmmo00}, as described in Methods, and are listed in Supplementary Table~1. {\bf c} Inter-cellular transcriptional activation of Engrailed by Wingless ($c$ = WGen). {\bf d} Intra-cellular cleavage of Cubitus Interruptus by Patched ($c$ = PTCCID). 

\newpage

\end{document}